\documentclass{article}
\usepackage[utf8]{inputenc}
\usepackage[T1]{fontenc}
\usepackage[english]{babel}
\usepackage{amssymb}
\usepackage{amsmath}
\usepackage{amsfonts} 
\usepackage{amssymb}
\usepackage{mathrsfs}
\usepackage{graphicx}
\usepackage{enumerate}
\usepackage{epstopdf}
\usepackage{caption}
\usepackage{subcaption} 	
\usepackage{float}
\usepackage{hyperref}

\usepackage{geometry}
\title{New $(2+1)$-dimensional Burgers equation and its solitary wave solutions via the Lie symmetry method.}

\author{{Nardjess Benoudina$^{1}$, Nassim Bessaad$^{2}$}\\
{\small \it\  $^1$ Department of Physics and Electronic Information
Engineering,} \\
{\small \it\ Zhejiang Normal University,} \\
{\small \it\ Jinhua, 321004, PR China }  \\
{\small \it\  $^2$School of Electronic Information and Electrical Engineering,} \\
{\small \it\ Shanghai Jiao Tong University, }\\
{\small \it\  Shanghai 200240, China} }

\begin{document}

\maketitle

\begin{abstract}
  In this paper, the new $(2+1)$-dimensional Burgers equation has been derived using the Burgers equation' recursion operator as follows
   \begin{equation*}
  u_{xt}+\left(u_{t}+uu_{x}-\nu u_{xx}\right)_{y}+3\left(u_{x}\partial_{x}^{-1}u_{y}\right)_{x}=0.
   \end{equation*}
   This nonlinear model is an interesting generalization of the Burgers equation. Because of its complexity, we have applied the Lie symmetry approach to an equivalent equation of the new Burgers equation to achieve 6-dimensional vector fields of symmetry. The reduction process under four symmetries subalgebras helps to investigate four simpler equations, one of which is the famous Riccati equation. Therefore, four explicit solutions are attained and graphically illustrated in $3$D and contour plots. Different solitary wave dynamics of the new $(2+1)$-dimensional Burgers equation are determined, which includes bright soliton, breather, kink, periodic solution and some interactions.
\end{abstract}


\section{Introduction}
The nonlinear Burgers equation is a fascinating model that appears in variety of fields, including aspects of turbulence, nonlinear wave propagation, shock wave theory, traffic flow, cosmology, molecular interface growth, gas dynamics, sedimentation of polydispersive suspensions and colloids, and longitudinal elastic waves in isotropic solids\cite{Bonkile2018,Nagatani2000}. In 1915, this model was first introduced by Harry Bateman \cite{Batman1915} as
\begin{equation}\label{Burgers}
  u_{t}+uu_{x}=\nu u_{xx},
\end{equation}
where $u(x,t)$ represents the velocity and $\nu$ is the kinematic viscosity. Later in 1948, Johannes Martinus Burger found a physical explanation of Eq.(\ref{Burgers}) in the theory of turbulence \cite{Burgers1948}. As a tribute to his work, this equation is commonly referred to as the "Burgers' equation". To more thoroughly describe nonlinear phenomena, new extensions of the latter are devised. In addition, their explicit and numerical solutions have been extensively investigated using different methods, such as the Riccati equation rational expansion method \cite{Wang2005}, tanh-coth method \cite{Wazwaz2007}, the ansatz method \cite{Jawad2010}, finite element method \cite{ARMINJON1978}, and others.
The construction of new generalized versions of Eq.\eqref{Burgers} is an important task \cite{Wazwaz2014a}. In this study. we have derived two Burgers equations using theory of constructing high-dimensional integrable equations \cite{Lou1997} and the Burgers' equation recursion operator. Then, some computation between the two equations lead us to determine the new $(2+1)$-dimensional Burgers equation 
\begin{equation}\label{newBurgers}
  u_{xt}+\left(u_{t}+uu_{x}-\nu u_{xx}\right)_{y}+3\left(u_{x}\partial_{x}^{-1}u_{y}\right)_{x}=0,
\end{equation}
where $u=u(x,y,t)$, and $\partial_{x}^{-1}=\left(\partial_{x}^{-1} g\right)(x)=\int_{-\infty}^{x} g(t) d t$ is the inverse operator. To avoid the integral term, we suggested the transformation $u(x,y,t)=w(x,y,t)_{x}$ to get simpler equivalent equation. The Lie symmetry approach \cite{Olver1993,Benoudina2020,Benoudina2022} is applied on the new equation to construct 6-dimensional vector fields of symmetry algebra. The equivalent equation of Eq.\eqref{newBurgers} is reduced under four subalgebras to get four explicit solutions.
As consequence, new four exact solutions of Eq.\eqref{newBurgers} has been deduced and graghically illustrated in $3$D and contour plots to show the solitary wave solutions, that are, bright soliton, breather, kink, periodic, and their interactions.

This study is organized as follows. The new $(2+1)$-dimensional Burgers equation \eqref{newBurgers} has been derived in Section~\ref{Bursec1}. The investigation of new solitary wave solutions of Eq.\eqref{newBurgers} via the Lie symmetry method in Section~\ref{Bursec2}. Finally, Section~\ref{Bursec3} summarizes this work in a conclusion.

\section{Derivation of new Burgers equation} \label{Bursec1}
This section aims to analyze the construction of a new integro-differential Burgers equation. Let us consider the Burgers equation's recursion operator
\begin{equation} \label{Burrecursion}
    \mathfrak{R} =-3u_{x}\partial_{x}^{-1}+u+\frac{1}{2}\nu\partial_{x}.
\end{equation}
We assume the following equation hierarchy based on high-dimensional integrable model construction via recursion operators theory in \cite{Lou1997,Benoudina2023}.
\begin{equation}
    u_{t}=\sum_{i=1}^{n} \alpha_{i} \mathfrak{R}^{s_{i}} u_{x_{i}},
    \end{equation} 
with $\alpha_{i}$ $(i=1,...,n)$ being arbitrary constants, and $s_{i}=0, \pm 1, \pm 2, \ldots$ for invertible $\mathfrak{R}(u)$ and $s_{i}=0, 1, 2, \ldots$ for non-invertible $\mathfrak{R}(u)$.
Next, the equation hierarchy for $\alpha_{i}=1, i=1,...,n$, $n=1$ and $s_{i}=1$ is written
\begin{equation} \label{Burrecu1}
    u_{t}= \mathfrak{R}(u_{x}),
  \end{equation}
where $u=u(x,y,t)$, and $\mathfrak{R}$ defines the Burgers' recursion operator \eqref{Burrecursion}. Hence, the Eq.\eqref{Burrecu1} becomes
\begin{equation} \label{Burg1}
    u_{t}+2uu_{x}-\frac{1}{2}\nu u_{xx}=0.
\end{equation}
Similarly, we consider
\begin{equation} \label{Burrecu2}
    u_{t}= \mathfrak{R}(u_{y}),
  \end{equation}
It results in the following equation
\begin{equation} \label{Burg2}
    u_{t}-uu_{y}-\frac{1}{2}\nu u_{xy}+3u_{x}\partial_{x}^{-1}u_{y}=0.
\end{equation}
Now, we derive Eqs.\eqref{Burg1}, \eqref{Burg2} with respect to $y$ and $x$, respectively. Then, we add the resulted equations by harnassing the property $u_{yxx}=u_{xxy}$ to find the new $(2+1)$-dimenional Burgers equation \eqref{newBurgers}.

\section{Lie Symmetry analysis and new invariant solutions of the $(2+1)$-dimensional Burgers equation} \label{Bursec2}

The utilization of the Lie symmetry technique proves to be a highly effective methodology in the resolution of NPDEs. Before applying this method, we transfer the equation \eqref{newBurgers} to simpler equation by assuming that $u(x,y,t)=w(x,y,t)_{x}$ which gives
\begin{equation}\label{newBurSimple}
    w_{xxt}+\left(w_{xt}+w_{x}w_{xx}-\nu w_{xxx}\right)_{y}+3\left(w_{xx}w_{y}\right)_{x}=0.
\end{equation}
We consider the 1-parameter Lie group of point transformations for the Eq.\eqref{newBurSimple}
\begin{equation}
\begin{array}{lllll}
  \breve{x}=x+\epsilon \xi_{1}(x, y, t, w)+O\left(\epsilon^{2}\right), \\
  \breve{y}=y+\epsilon \xi_{2}(x, y, t, w)+O\left(\epsilon^{2}\right), \\
   \breve{t}=t+\epsilon \xi_{3}(x, y, t, w)+O\left(\epsilon^{2}\right), \\
  \breve{w}=w+\epsilon \eta(x, y, t, w)+O\left(\epsilon^{2}\right),
\end{array}
\end{equation}
where $\epsilon $ is a small parameter. The determined infinitesimals $(\xi_{1},\, \xi_{2}, \, \xi_{3}, \eta)(x, y, t, w)$ state the infinitesimal generator $\mathbb{V}$ as 

\begin{equation}
\mathbb{V}=\xi_{1} \frac{\partial}{\partial x}+\xi_{2} \frac{\partial}{\partial y}+\xi_{3}\frac{\partial}{\partial t}+\eta \frac{\partial}{\partial w}.
\end{equation}
Then, the infinitesimals can be generated by satisfying the invariance condition $\left.p r^{(4)}   \mathbb{V}(\Delta)\right|_{\Delta=0}=0$, where $pr^{(4)} \mathbb{V}$ is the forth prolongation of $\mathbb{V}$ and $\Delta= w_{xxt}+\left(w_{xt}+w_{x}w_{xx}-\nu w_{xxx}\right)_{y}+3\left(w_{xx}w_{y}\right)_{x}$. Therefore, we get the following system of determined equations using Maple :

\begin{equation}
  \begin{aligned}
      &\left(\eta\right)_w=0,\left(\eta\right)_x=\left(\xi_1\right)_t,\left(\eta\right)_y=\frac{\left(\xi_1\right)_t}{3},\left(\xi_3\right)_t=2\left(\xi_2\right)_y,\left(\xi_3\right)_w=0,\left(\xi_3\right)_x=0,\left(\xi_3\right)_y=0 \\
      &\left(\xi_1\right)_w=0,\left(\xi_1\right)_x=\left(\xi_2\right)_y,\left(\xi_1\right)_y=0,\left(\xi_2\right)_t=0,\left(\xi_2\right)_w=0,\left(\xi_1\right)_x=0,\left(\xi_2\right)_{y, y}=0.
       \end{aligned}
\end{equation}
  When we solve the above system, we have the infinitesimals.
  
  \begin{equation}
    \xi_x=a_{1} x+F 1(t),\quad \xi_y=z_{1} y+a_{2}, \quad \xi_t=2a_{1} t+a_{3},\quad \eta_w=\frac{(3 x+y)F_{1}(t)}{3}+F_{2}(t).
    \end{equation}
Setting $F_{1}(t)=a_{4}t+a_{5}$, and $F_{2}(t)=a{6}t$. Then, we collect the coefficients to determine the symmetries 
\begin{equation} \label{BurgSym}
 \mathbb{V}_{1} = \frac{\partial}{\partial x}, \quad
 \mathbb{V}_{2} = \frac{\partial}{\partial y} , \quad
 \mathbb{V}_{3} = \frac{\partial}{\partial t} ,\quad
 \mathbb{V}_{4} = t\frac{\partial}{\partial w} ,\quad
 \mathbb{V}_{5} =x\frac{\partial}{\partial x}+y\frac{\partial}{\partial y}+2t\frac{\partial}{\partial t},\quad
 \mathbb{V}_{6} = t\frac{\partial}{\partial x}+\left(x+\frac{1}{3}y\right)\frac{\partial}{\partial w}.
\end{equation}

As a result, 6-dimensional vector fields of symmetry algebra $\mathfrak{G}$ spanned by $\{\mathbb{V}_{1},...,\mathbb{V}_{6}\}$ is constructed.

\subsection{Reduction of Eq.\eqref{newBurSimple} and new invariant solutions of Eq.\eqref{newBurgers}}

\subsubsection{Subalgebra : $\mathbb{V}_{2}=\frac{\partial}{\partial{y}}$}
The characteristic equation of the symmetry subalgebra $\mathbb{V}_{2}$ is 
\begin{equation} \label{BurgCharact}
    \frac{dx}{0}=\frac{dy}{1}=\frac{dt}{0}=\frac{dw}{0}.
    \end{equation} 
By integrating Eqs.\eqref{BurgCharact}, we find the group-invariant function $ w(x,y,t)=f(\xi,\zeta)$ and the similarity variables $\xi=x$, and $\zeta=y$. Substituting into Eq.\eqref{newBurSimple} to find the simple equation $f_{\zeta \xi \xi}=0$, by triple integration of the latter and the invariants,
we have 
\begin{equation}\label{Bsolsim1}
   w(x,y,t)= F_{3}(x)+F_{2}(t)+F_{1}(t) x.
\end{equation}
Hence, the Eq.\eqref{newBurgers} has the following solution

\begin{equation}\label{solburg1}
    w(x,y,t)= F_{3}'(x)+F_{1}(t) x,
 \end{equation}
where $F_{1}(t)$, $F_{2}(t)$, and $F_{3}(x)$ are arbitrary constants

\begin{figure}
    \centering
      \begin{subfigure}[b]{0.24\textwidth}
        \centering
          \includegraphics[width=\textwidth]{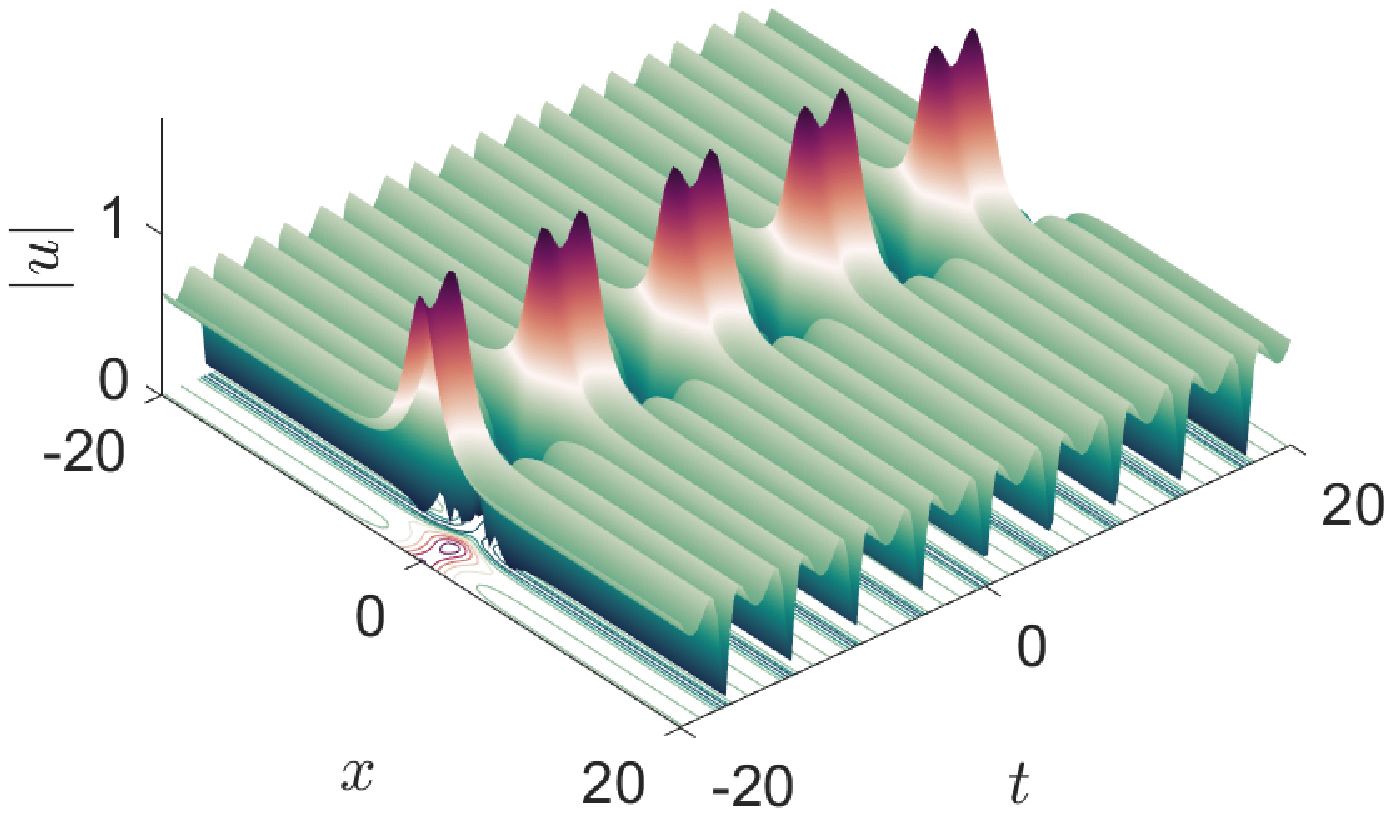}
          \caption{}
          \label{bfig1_1}
      \end{subfigure}
      \centering
      \begin{subfigure}[b]{0.2\textwidth}
        \centering
          \includegraphics[width=\textwidth]{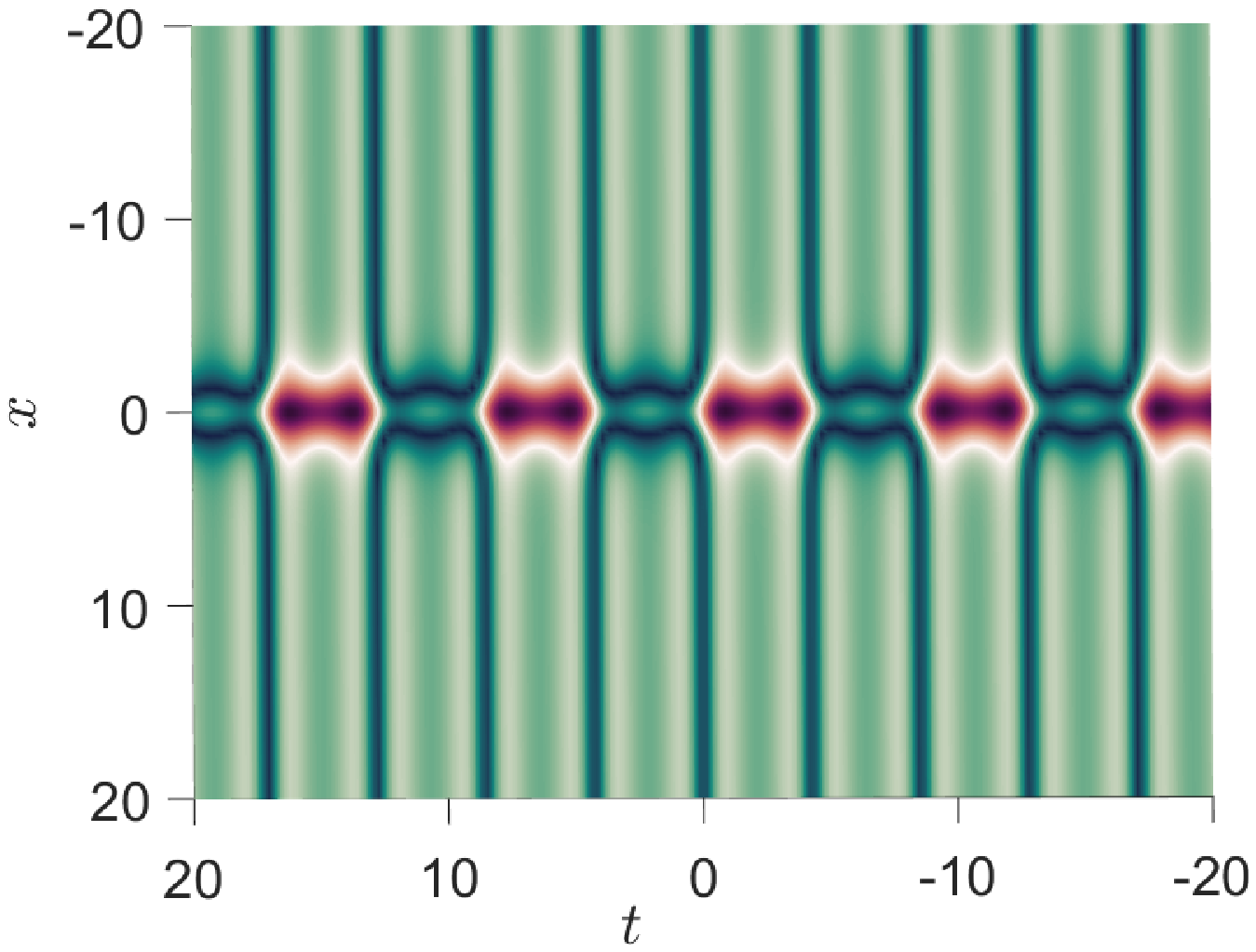}
          \caption{}
          \label{bfig1_1v}
      \end{subfigure}
      \centering
      \begin{subfigure}[b]{0.24\textwidth}
        \centering
          \includegraphics[width=\textwidth]{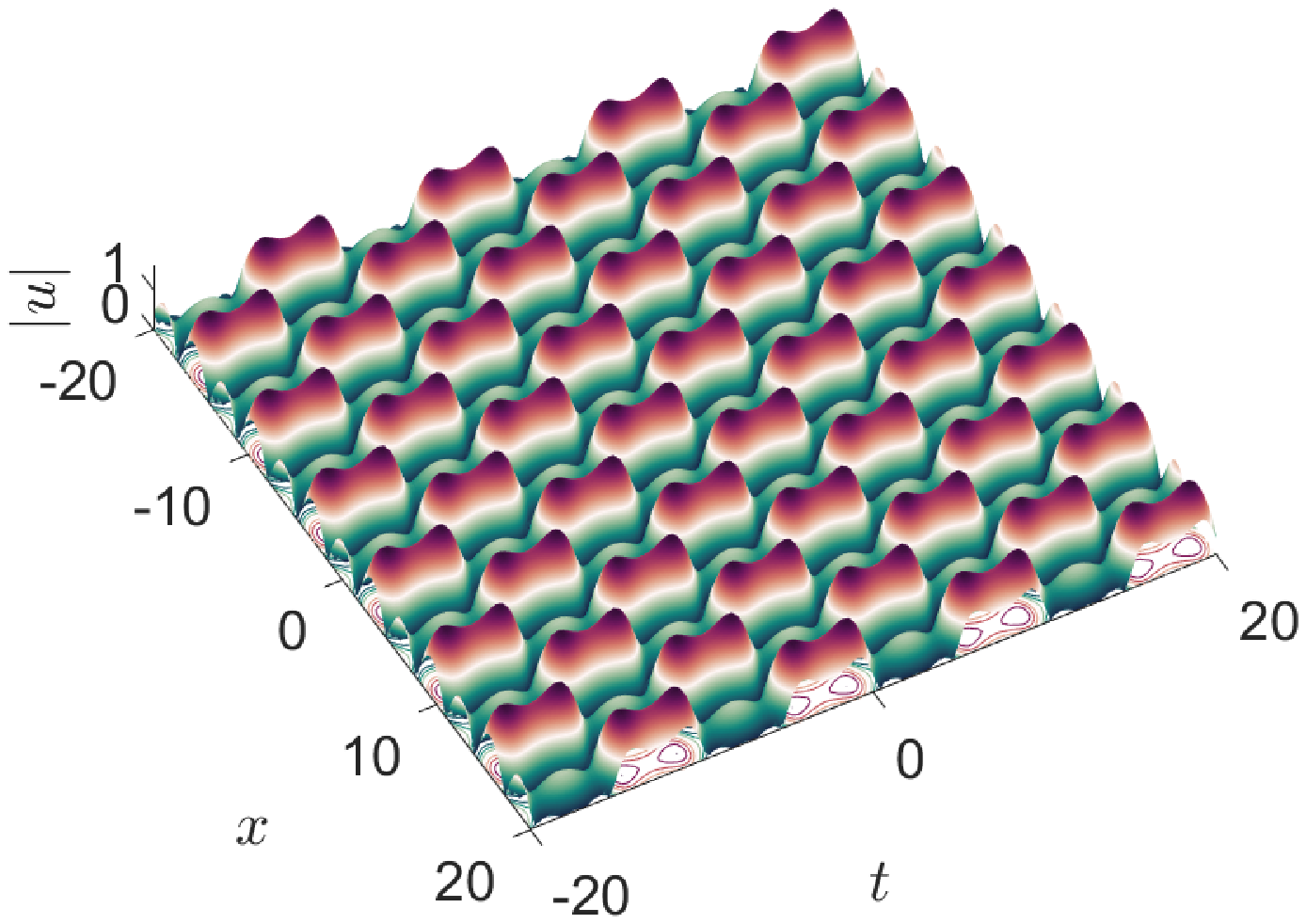}
          \caption{}
          \label{bfig1_2}
      \end{subfigure}
      \centering
      \begin{subfigure}[b]{0.2\textwidth}
        \centering
          \includegraphics[width=\textwidth]{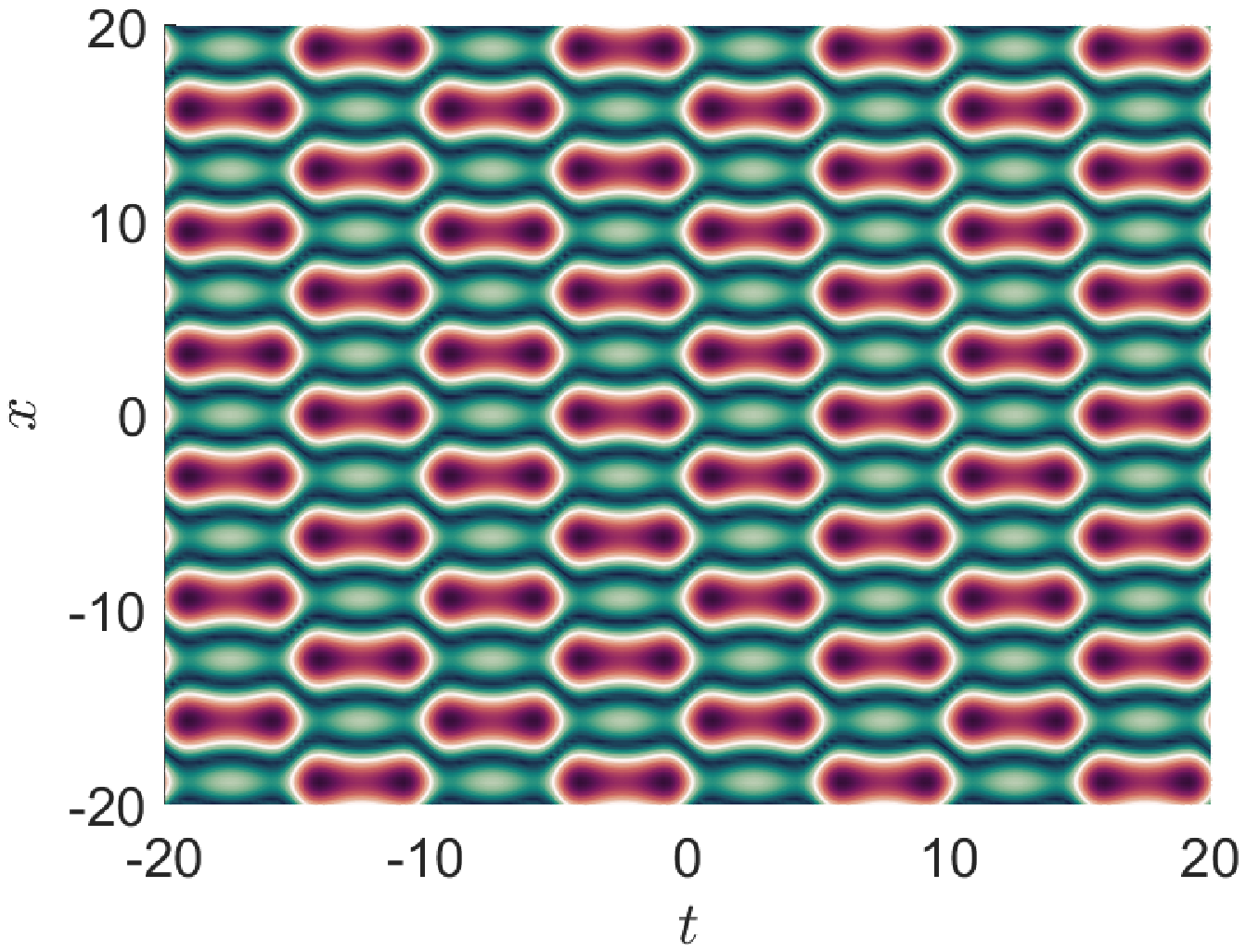}
          \caption{}
          \label{bfig1_2v}
      \end{subfigure}
      \centering
      \begin{subfigure}[b]{0.24\textwidth}
        \centering
          \includegraphics[width=\textwidth]{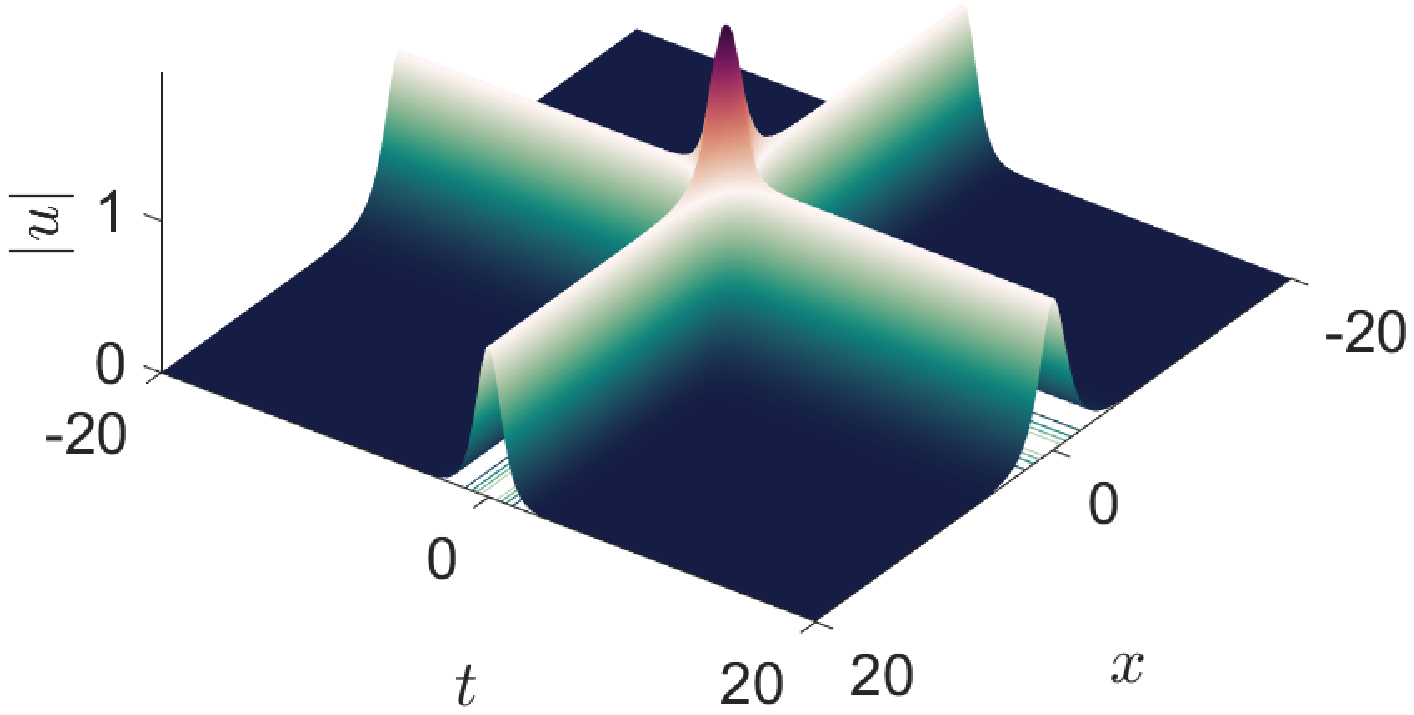}
          \caption{}
          \label{bfig1_3}
      \end{subfigure}
      \centering
      \begin{subfigure}[b]{0.2\textwidth}
        \centering
          \includegraphics[width=\textwidth]{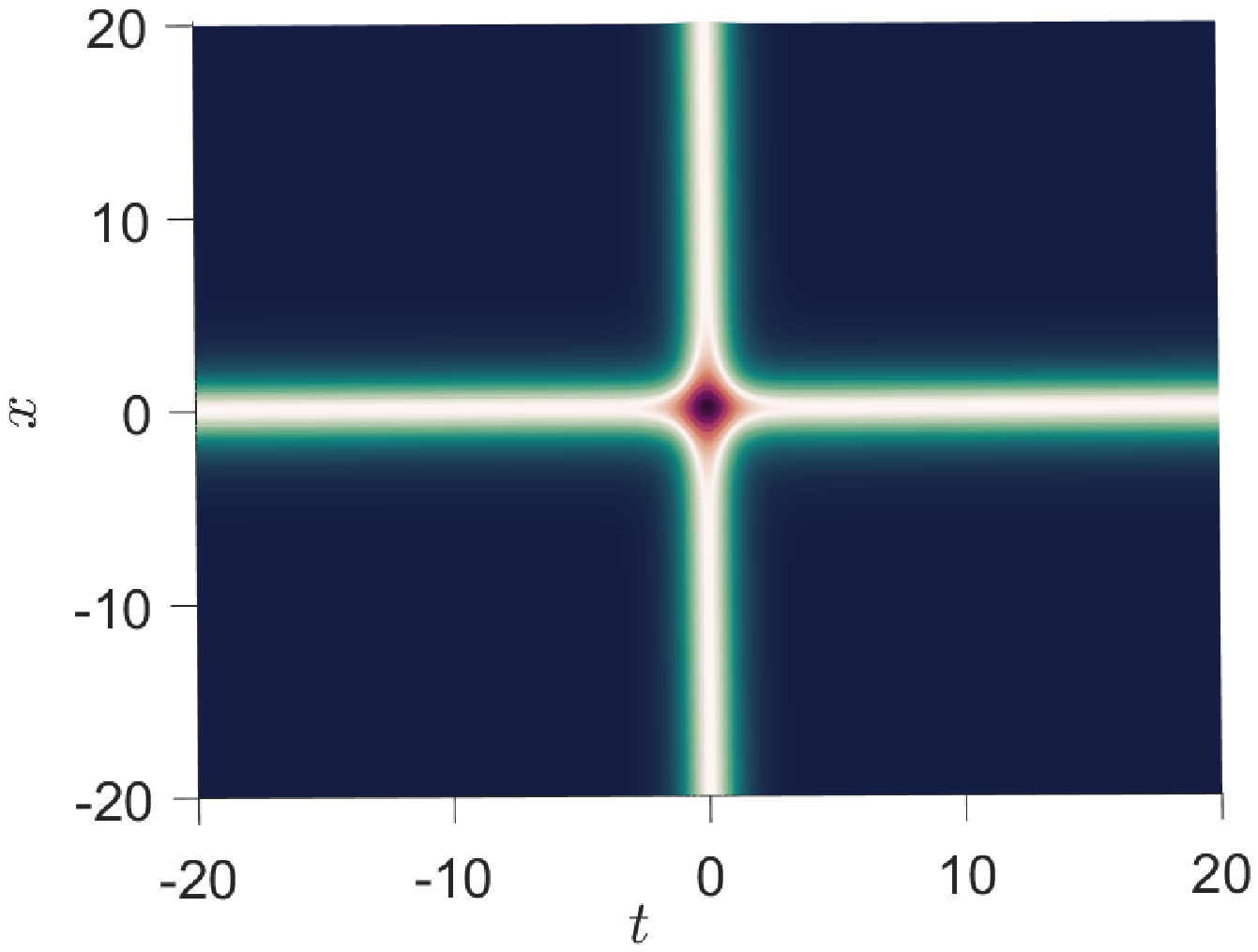}
          \caption{}
          \label{bfig1_3v}
      \end{subfigure}
      \caption{Profiles of the dynamical behavior of the solution \eqref{solburg1} in 3D and contour plots. Figures \ref{bfig1_1},\ref{bfig1_1v}: breather soliton on periodic bachkground for $F_{1}(t)=\operatorname{sech}(t)$, $F_{3}(x)=\operatorname{dn}(x,1.2)$. Figures \ref{bfig1_2}, \ref{bfig1_2v}: periodic breather for $F_{1}(t)=\cos(t)$, $F_{3}(x)=\operatorname{dn}(x,1.1)$. Figures \ref{bfig1_3}, \ref{bfig1_3v}: Two bright soliton interaction for $F_{1}(t)=\operatorname{sech}(t)$, $F_{3}(x)=\tanh(x)$.}\label{Burfig1}
    \end{figure}

    \subsubsection{Subalgebra: $\mathbb{V}_{3}=\frac{\partial}{\partial{t}}$}
    In this case, The symmetry $\mathbb{V}_{3}$ has the similarity variables $\xi=x$, $\zeta=y$, and the group-invariant form is $w(x,y,t)=f(\xi,\zeta)$, substituting into Eq.\eqref{newBurSimple} to investigate the reduced equation
    \begin{equation}\label{Breduced1}
        -\nu f_{\xi, \xi, \xi, \zeta}+4 f_{\xi, \zeta} f_{\xi, \xi}+f_\xi f_{\xi, \xi, \zeta}+3 f_{\xi, \xi, \xi} f_\zeta=0.
    \end{equation}
We reapply the Lie symmetry again to get the ODE $ \nu g_{\rho, \rho, \rho, \rho} -4 g_{\rho, \rho}^2-4 g_{\rho} g_{\rho, \rho, \rho}=0$, where $\rho=-y \sin (y)-\cos (y)+x$ and $g(\rho)=f(\xi,\zeta)$. We integrate the ODE twice to find $\nu g_{\rho, \rho}-2 g_{\rho}^2=0$, then solve it by Maple to attain new 
solution of Eq.\eqref{newBurSimple}
\begin{equation}\label{Bsolsim2}
    w(x, y, t)=-\frac{1}{2}v\left(\ln (2)+\ln \left(\frac{\alpha_{1}\left(\sin (y) y+\cos (y)- x\right)-\alpha_{2}}{v}\right)\right).
\end{equation}
From the solution \eqref{Bsolsim2}, we conclude the second solution of Eq.\eqref{newBurgers}

\begin{equation}\label{solBurg2}
    u(x, y, t)=\frac{\nu \alpha_{1}}{2 \alpha_{1}\left(y \sin (y)+\cos (y)-x\right)-2\alpha_{2}},
\end{equation}
and $\alpha_{1}$, $\alpha_{2}$ being arbitrary constants.
\begin{figure}[H]
    \centering
      \begin{subfigure}[b]{0.4\textwidth}
        \centering
          \includegraphics[width=\textwidth]{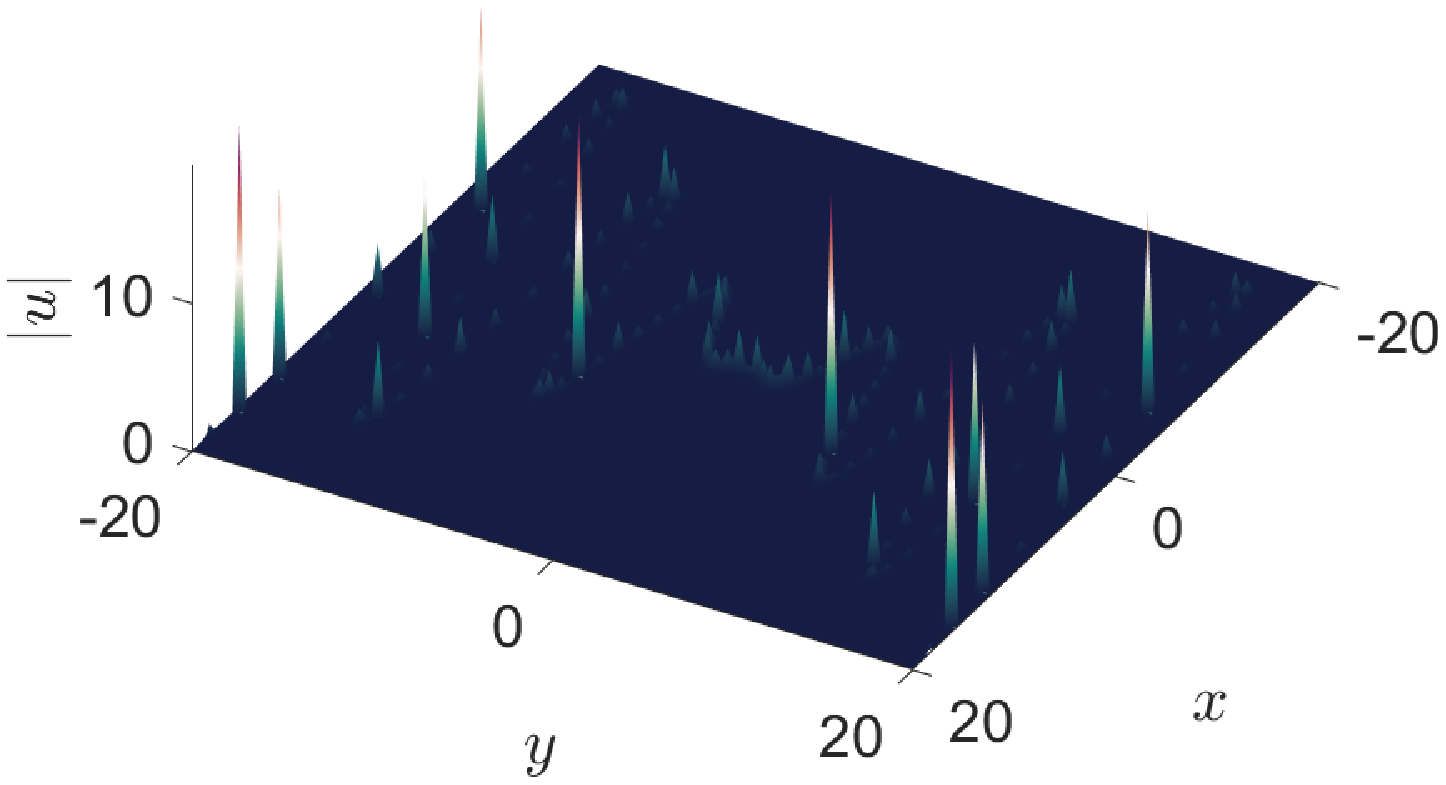}
          \caption{}
          \label{bfig2}
      \end{subfigure}
      \centering
      \begin{subfigure}[b]{0.4\textwidth}
        \centering
          \includegraphics[width=\textwidth]{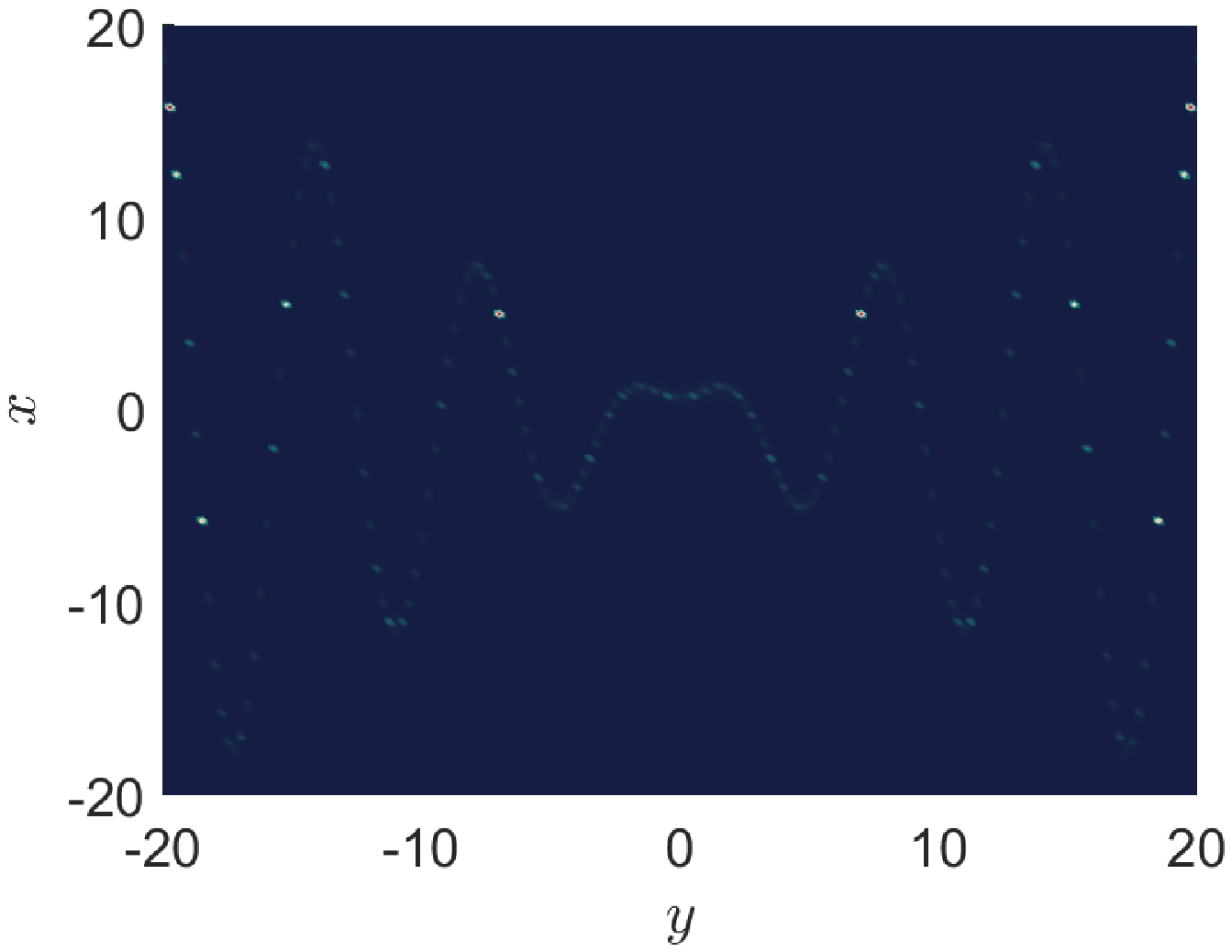}
          \caption{}
          \label{bfig2}
      \end{subfigure}
      \caption{Breather soliton with periodic moving path of the solution \eqref{solBurg2} for the parameters $\alpha_{1}=0.576$, $\alpha_{2}=1$, $\nu=0.8$.}\label{Burfig1}
    \end{figure}

\subsubsection{Subalgebra: $\mathbb{V}_{7}=b_{1}\mathbb{V}_{2}+c_{1}\mathbb{V}_{3}=b_{1}\frac{\partial}{\partial{y}}+c_{1}\frac{\partial}{\partial{t}}$}
By employing the subalgebra $\mathbb{V}_{7}$, the group-invariant solutions $w(x,y,t)=f(\xi,\zeta)$ has been achieved, and the similarity variables $\xi=x$, and $\zeta=\frac{ b_1t-c_1 y}{b_1}$ which lead to 
\begin{equation}\label{Breduced2}
    \nu f_{\xi, \xi, \xi, \zeta} c_1+\left(-c_1 f_\xi+b_1\right) f_{\xi, \xi, \zeta}-4 c_1\left(f_{\xi, \zeta} f_{\xi, \xi}+\frac{3 f_{\xi, \xi, \xi} f_{\zeta}}{4}+\frac{f_{\xi, \zeta, \zeta}}{4}\right)=0.
\end{equation}
We solve the Eq.\eqref{Breduced2} using Maple, by the invariants, we get 
\begin{equation}\label{Bsolsim3}
    w(x, y, t)=-\frac{1}{2} \nu \ln \left(\left(\alpha_{2}(x-y)+\alpha_{1}\right) +\frac{\alpha_{2}  b_1 t}{c_1}\right)+\alpha_{3}.
\end{equation}
Therefore, the solution of Eq.\eqref{newBurgers} is 
\begin{equation}\label{solBurg3}
    u(x, y, t)=-\frac{ \nu \alpha_{2} \alpha_1}{2 \alpha_1 \left(( x- y)\alpha_{2}+\alpha_{1}\right) +2\alpha_{2}  b_1 t},
\end{equation}
where $\alpha_{1}$, $\alpha_{2}$, $b_{1}$ and $c_{1}$ are arbitrary constants.

\begin{figure}[H]
    \centering
      \begin{subfigure}[b]{0.4\textwidth}
        \centering
          \includegraphics[width=\textwidth]{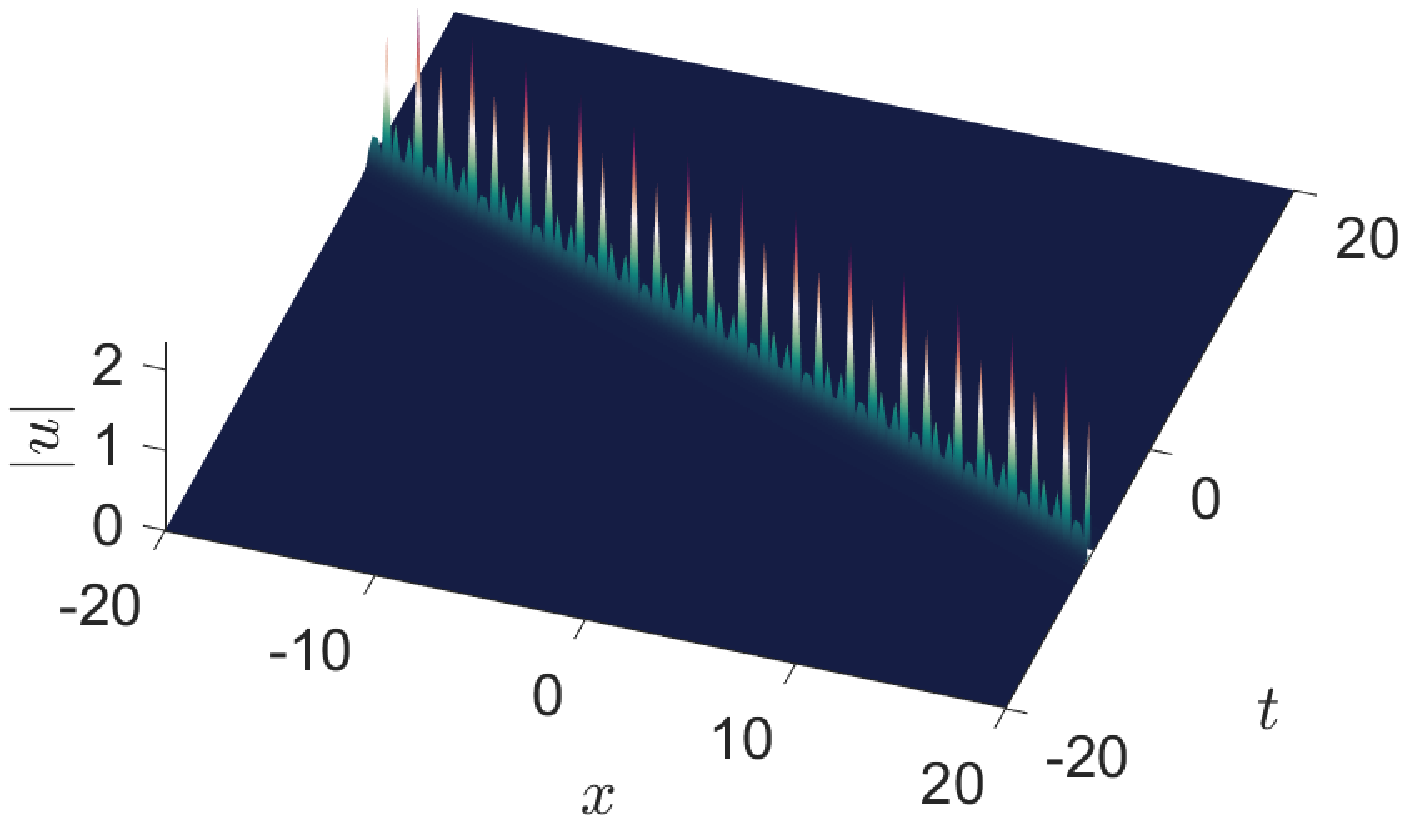}
          \caption{}
          \label{bfi3}
      \end{subfigure}
      \centering
      \begin{subfigure}[b]{0.4\textwidth}
        \centering
          \includegraphics[width=\textwidth]{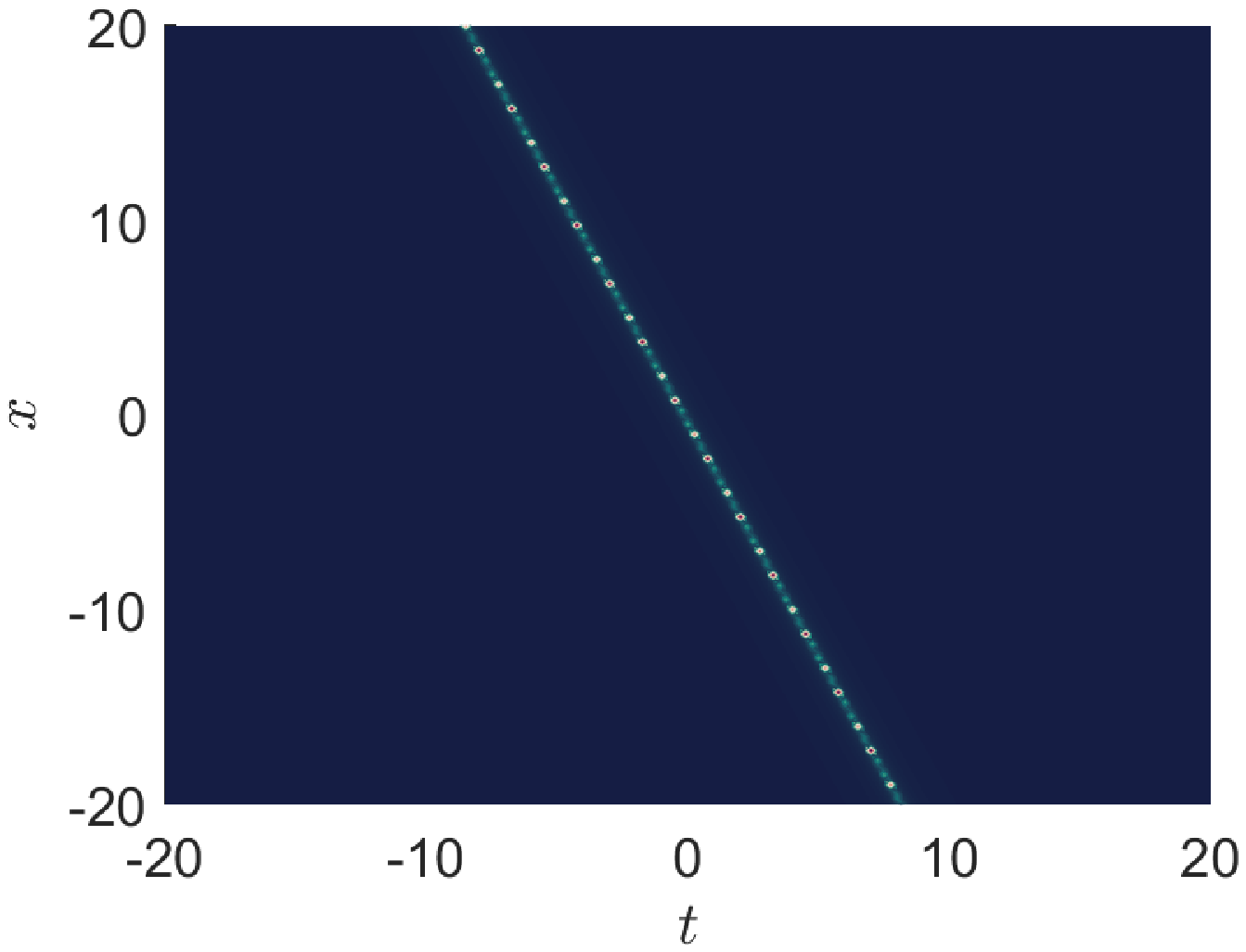}
          \caption{}
          \label{bfig3v}
      \end{subfigure}
\caption{The dynamical behavior of the solution \eqref{solBurg3} shows a breather for $c_{1}=0.2$, $b_{1}=0$, $\alpha_{1}=1$, $\alpha_{2}=0.7$, $\nu=0.1$, and $y=1$.}\label{Burfig1}
\end{figure}

\subsubsection{Subalgebra: $\mathbb{V}_{8}=\mathbb{V}_{1}+\mathbb{V}_{2}+\mathbb{V}_{3}=\frac{\partial}{\partial{x}}+\frac{\partial}{\partial{y}}+\frac{\partial}{\partial{t}}$}
The similarity variables of the symmetry $\mathbb{V}_{8}$ are $X=y-x$, $Y=t-x$, and its invariant form $w(x,y,t)=f(X,Y)$, substituting into Eq.\eqref{newBurSimple} to reach 
\begin{equation}\label{Breduced3}
    \begin{aligned}
    &\nu \left(f_{XXXX}+3  f_{XXXY}+3 f_{XXYY}+ f_{XYYY}\right) +\left(-10 f_X-f_Y+1\right) f_{XYY}+\left(-4 f_X-f_Y\right)f_{XXX} \\
    &  +\left(-11 f_X-2 f_Y\right) f_{XXY}+\left(-3 f_X+1\right) f_{YYY}-4\left(f_{XX}+f_{XY}\right)\left(f_{XX}+2 f_{XY}+f_{YY}\right)=0.
    \end{aligned}
\end{equation}
We reduce the Eq.\eqref{Breduced3} using the Lie symmetry again to get the ODE
\begin{equation}\label{BODE2}
    -32 g_{\xi \xi}^2-32 g_\xi g_{\xi \xi \xi}+8 \nu g_{\xi \xi \xi \xi}+2 g_{\xi \xi \xi}=0,
\end{equation}
where $\xi=X+Y$, and $g(\xi)=f(X,Y)$. Integrating the ODE \eqref{BODE2} twice to achieve $8 \nu g_{\xi \xi}-16 g_\xi^2+2 g_\xi=0$. Then, substituting $g_{\xi}=h(\xi)$ into the latter ODE, we get the famous Riccati equation $ 8 \nu h_\xi-16 h(\xi)^2+2 h(\xi)=0$. As consequence, by adopting the Riccati equation solution and the invariants,
we investigate 
\begin{equation}\label{Bsolsim4}
    w(x, y, t)=\frac{-2 x+y+t}{8}-\frac{\nu \ln \left(8+\alpha e^{\frac{-2 x+y+t}{4 \nu}}\right)}{2}.
    \end{equation}

    Thus, the last solution of the new Burgers \eqref{newBurgers} is found as 

    \begin{equation}\label{solBurg4}
        u(x, y, t)=-\frac{2}{8+\alpha\mathrm{e}^{\frac{-2 x+y+t}{4 \nu}}},
     \end{equation}
where $\alpha$ being arbitrary constant.

     \begin{figure}[H]
        \centering
          \begin{subfigure}[b]{0.4\textwidth}
            \centering
              \includegraphics[width=\textwidth]{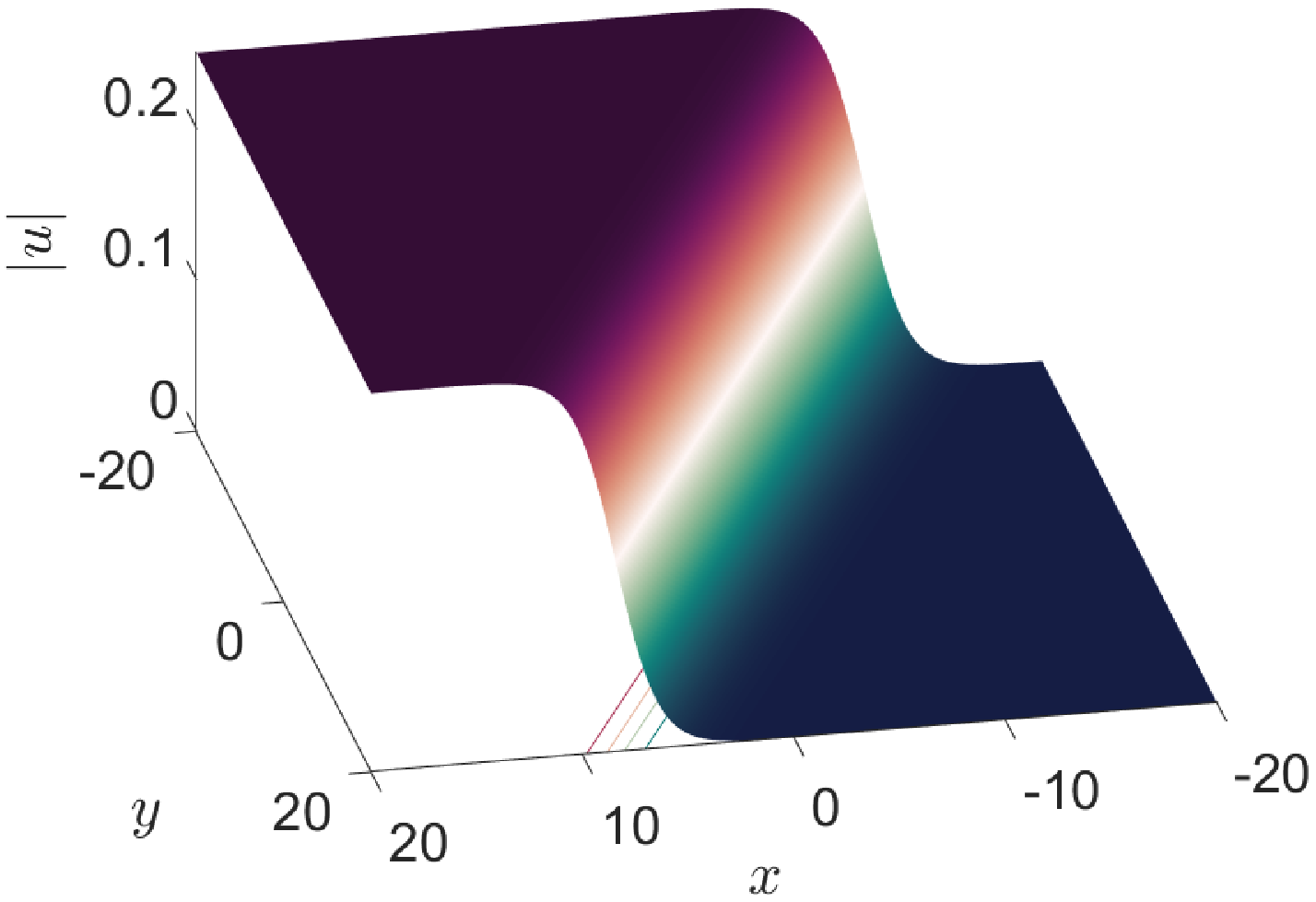}
              \caption{}
              \label{bfi4}
          \end{subfigure}
          \centering
          \begin{subfigure}[b]{0.4\textwidth}
            \centering
              \includegraphics[width=\textwidth]{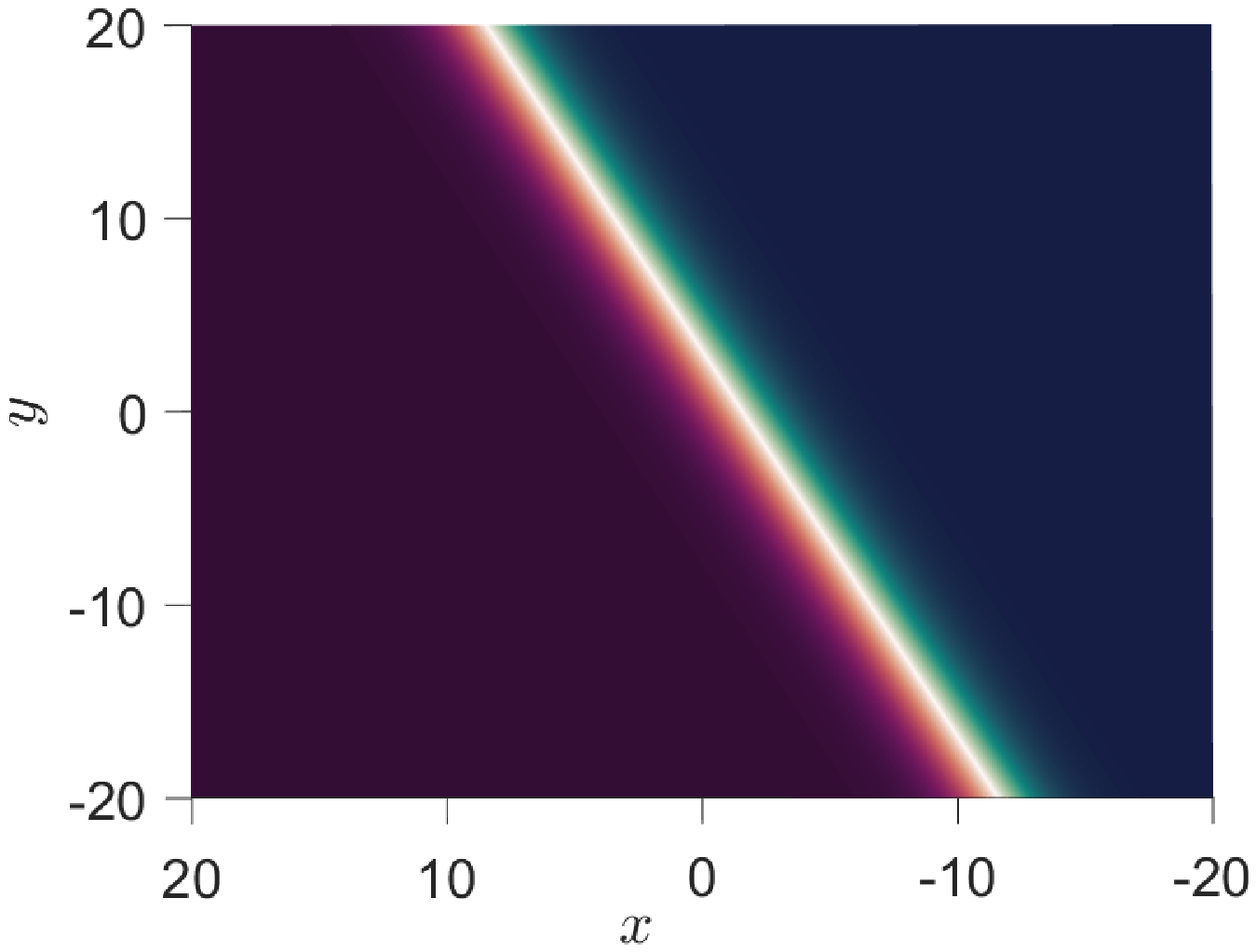}
              \caption{}
              \label{bfig4v}
          \end{subfigure}
    \caption{The dynamical behavior of the solution \eqref{solBurg4} reveals a kink for the parameters $\alpha=1$, $\nu=0.5$, $t=1$.}\label{Burfig1}
    \end{figure}

\section{Conclusion} \label{Bursec3}
Many notable results have been achieved in this study, which appeared as follows. First, we aim to derive a novel $(2+1)$-dimensional Burgers equation by employing the recursion operator and implementing a series of computation-guided procedures. Second, the utilization of the Lie symmetry method on the new Burgers equation's equivalent model has proven to be effective in the process of simplifying and examining four exact solutions of the Burgers equation. Finally, different solitary wave behaviors are determined for the new equation, which are, bright soliton, breather, periodic, kink, and their interactions. This intriguing model will contribute to developing our understanding of several physical phenomena.

\bibliographystyle{ieeetr}
\bibliography{Burgers}

\begin{thebibliography}{10}

\bibitem{Bonkile2018}
M.~P. Bonkile, A.~Awasthi, C.~Lakshmi, V.~Mukundan, and V.~S. Aswin, ``{A
  systematic literature review of Burgers' equation with recent advances},''
  {\em Pramana - J. Phys.}, vol.~90, no.~6, pp.~1--21, 2018.

\bibitem{Nagatani2000}
T.~Nagatani, ``{Density waves in traffic flow},'' {\em Phys. Rev. E - Stat.
  Physics, Plasmas, Fluids, Relat. Interdiscip. Top.}, vol.~61, no.~4,
  pp.~3564--3570, 2000.

\bibitem{Batman1915}
H.~Batman, ``{Some recent researches on the motion of fluids},'' vol.~43,
  no.~4, pp.~163--170, 1915.

\bibitem{Burgers1948}
J.~M. Burgers, ``{A mathematical model illustrating the theory of
  turbulence},'' {\em Adv. Appl. Mech.}, vol.~1, no.~C, pp.~171--199, 1948.

\bibitem{Wang2005}
Q.~Wang, Y.~Chen, and H.~Zhang, ``{A new Riccati equation rational expansion
  method and its application to (2 + 1)-dimensional Burgers equation},'' {\em
  Chaos, Solitons and Fractals}, vol.~25, no.~5, pp.~1019--1028, 2005.

\bibitem{Wazwaz2007}
A.~M. Wazwaz, ``{Multiple-front solutions for the Burgers equation and the
  coupled Burgers equations},'' {\em Appl. Math. Comput.}, vol.~190, no.~2,
  pp.~1198--1206, 2007.

\bibitem{Jawad2010}
A.~J. a.~M. Jawad, M.~D. Petkovi{\'{c}}, and A.~Biswas, ``{Soliton solutions of
  Burgers equations and perturbed Burgers equation},'' {\em Appl. Math.
  Comput.}, vol.~216, no.~11, pp.~3370--3377, 2010.

\bibitem{ARMINJON1978}
P.~Arminjon and C.~Beauchamp, ``{A finite element method for Burgers' equation
  in hydrodynamics},'' {\em Int. J. Numer. Methods Eng.}, vol.~12, no.~3,
  pp.~415--428, 1978.

\bibitem{Wazwaz2014a}
A.~M. Wazwaz, ``{A study on a (2 + 1)-dimensional and a (3 + 1)-dimensional
  generalized Burgers equation},'' {\em Appl. Math. Lett.}, vol.~31,
  pp.~41--45, 2014.

\bibitem{Lou1997}
S.~Lou, ``{Higher-dimensional integrable models with a common recursion
  operator},'' {\em Commun. Theor. Phys.}, vol.~28, no.~1, pp.~41--50, 1997.

\bibitem{Olver1993}
P.~J. Olver, {\em {Applications of Lie Groups to Differential Equations}}.
\newblock New York: Springer, 1993.

\bibitem{Benoudina2020}
N.~Benoudina, Y.~Zhang, and C.~M. Khalique, ``{Lie symmetry analysis, optimal
  system, new solitary wave solutions and conservation laws of the Pavlov
  equation},'' {\em Commun. Nonlinear Sci. Numer. Simul.}, vol.~94, p.~105560,
  2021.

\bibitem{Benoudina2022}
N.~Benoudina, Y.~Zhang, C.~M. Khalique, and N.~Bessaad, ``{Novel hybrid
  solitary waves and shrunken-period solutions, solitary Moir{\'{e}} pattern
  and conserved vectors of the (4 + 1)-Fokas equation},'' {\em Int. J. Geom.
  Methods Mod. Phys.}, vol.~19, no.~12, p.~2250195, 2022.

\bibitem{Benoudina2023}
N.~Benoudina, Y.~Zhang, and N.~Bessaad, ``{A new derivation of ( 2 + 1
  )-dimensional Schr{\"{o}}dinger equation with separated real and imaginary
  parts of the},'' {\em Nonlinear Dyn.}, vol.~111, no.~7, pp.~6711--6726, 2023.

\end{thebibliography}

\end{document}